\documentclass[letterpaper, 10 pt, conference]{ieeeconf}  

\IEEEoverridecommandlockouts                              

\usepackage{graphics} 
\usepackage{epsfig} 
\usepackage{graphicx}
\usepackage{cite}
\usepackage{makecell}
\usepackage{url}
\usepackage{xcolor}
\usepackage{booktabs}
\usepackage{array}
\usepackage{tabularx}
\newcolumntype{Y}{>{\centering\arraybackslash}X}

\newcommand{\Tstrut}{\rule{0pt}{3.2ex}}        
\newcommand{\Bstrut}{\rule[-1.4ex]{0pt}{0pt}}  

\usepackage{amsmath} 
\usepackage{amssymb}  
\title{\LARGE \bf
A Two Stage Pipeline for Left Atrial Wall Constrained Scar Segmentation and Localization from LGE-MR Images}
\author{Bipasha Kundu$^{1}$and Cristian Linte$^{1,2}$
\thanks{$^{1}$Bipasha Kundu is with the Center for Imaging Science, Rochester Institute of Technology
Rochester, NY 14623, USA.  {\tt\small bk7944@g.rit.edu}}%
\thanks{$^{1,2}$Cristian Linte is with the Department of Biomedical Engineering and the Center for Imaging Science, Rochester Institute of Technology, USA, NY
14623, USA. {\tt\small calbme@rit.edu}}
}

\begin{document}

\maketitle
\thispagestyle{empty}
\pagestyle{empty}

\begin{abstract}
Accurate segmentation and localization of left atrial (LA) ablation scars from Late gadolinium enhancement (LGE)-MRI is essential for assessing the lesion completeness and guiding ablation therapy. Incomplete or discontinuous lesions can increase the recurrence rate of the therapy and inaccurate localization can misguide treatment planning. However, reliable quantification and localization of scar in LGE-MRI is challenging. The severely class imbalanced scar voxels, thin structure of the LA wall, and weak tissue contrast often lead to unrealistic scar predictions. In this paper, we propose a two stage nnUNet based framework  that takes LA anatomy into account to help with more precise scar localization and segmentation. In the first stage, an nnUNet model is trained to segment the LA cavity. In the second stage, patient specific cavity and wall signed distance maps (SDMs) are derived from the predicted anatomy to use as geometry aware inputs, and explicitly encode each voxel's signed spatial relationship to the atrial cavity and wall. This approach transforms scar segmentation from a solely intensity-based classification into anatomy-conditioned localization task, providing a continuous spatial prior that stabilizes learning for the thin atrial wall and suppresses topologically invalid predictions. To further address boundary ambiguity, we introduce a wall ROI-masked weighted loss combined with boundary uncertainty-aware supervision strategy that restricts learning to the atrial wall, while accounting for severe class imbalance. We evaluated our approach on the LAScarQS 2022 dataset and achieved a Dice of 61.1\% and ASSD of 1.711mm. Our reliable and effective framework improves scar segmentation and localization accuracy by enforcing anatomical validity through geometry-aware supervision, and lowering the false positive detections far away from the atrial wall.
\newline

\end{abstract}

\section{INTRODUCTION}


Atrial fibrillation (AFib), the most common sustained cardiac arrhythmia, affects an estimated 10.5 million adults in the United States and is nearly three times more prevalent than previously reported~\cite{AbbieJordan}. AFib originates in the upper chambers of the heart and is commonly treated using radio-frequency catheter ablation, in which targeted myocardial tissue is intentionally scarred to interrupt abnormal electrical conduction. However, 20\% - 40\% post-ablation recurrence often occurs, and patients are required to undergo multiple repeat procedures, increasing procedural risk and healthcare burden\cite{recur-rate}. Accurate localization and segmentation of LA scar tissue region enable patient specific ablation, lesion placement, reducing the procedural risk and recurrence rate to provide long term treatment success and reduce unnecessary tissue damage~\cite{Marrouche2014}. 

Late gadolinium enhancement magnetic resonance imaging (LGE-MRI) has been increasingly used to characterize and visualize left atrial (LA) anatomy and ablation-induced scar tissue. Anatomically, atrial scar resulting from ablation therapy is expected to lie on the thin myocardial wall of the LA\cite{zhou2022edge}. However, scar mapping is difficult due to the weak contrast, extremely thin wall (1-2 voxels), and severely class imbalanced (scar is $<$ 5\% of the LA wall). In addition, the manual annotation is extensively time consuming and subjective. The automated pixel-wise learning segmentation on the entire image often biases the model towards the background predictions and hallucinates the presence of scar in the blood pool or surrounding myocardium. As a result, it predicts false positives (FPs) with standard segmentation loss and  misses wall confined fibrosis by producing clinically unreliable scar maps.

With the rapid advancement of deep learning in medical imaging applications such as disease detection, classification or segmentation, increased attention have been directed towards the automated LA cavity and scar segmentation from LGE-MRI. While LA cavity segmentation has been extensively studied with strong performance reported across multiple datasets\cite{kundu2025investigating,kundu2025multi}, accurate segmentation of atrial scar remains comparatively underexplored and more challenging. For LGE-MRI in particular, hyper intense signal is not specific to fibrosis, as residual blood pool enhancement, partial volume effects in the thin atrial wall, and surrounding structures can also exhibit similar intensities that leads the intensity driven models to misclassify non-scar regions as scar. Yang \textit{et al.}\cite{yang2017fully} introduced one of the earliest deep learning approaches for LA scar segmentation, using super pixel based feature extraction and stacked sparse autoencoders. Khan \textit{et al.}\cite{khan2022sequential} implemented a two stage method and used boundary2patch to segment scar around the LA cavity. Zhang \textit{et al.}\cite{zhang2024left} used Convolutional Block Attention Module and edge attention module with residual nnUNet. Many studies have explored joint segmentation and quantification of atrial scar in concert with LA segmentation. To mitigate the dominance of background regions and emphasize anatomically relevant structures, LGE-MRI volumes have been center cropped to restrict analysis to a ROI surrounding the LA\cite{zhou2022edge, li2022atrialjsqnet,lefebvre2022lassnet}, but the reported performance is significantly poor due to its inherited challenges mentioned earlier. Using the same dataset as the challenge described in \cite{Lees-Miller-LaTeX-course-1}, prior methods report the mean scar dice on test dataset ranged from 0.47 to 0.59, 
demonstrating the persistent difficulties associated with sufficiently quantification of scar.

These limitations suggest that architectural changes or pixel-wise supervision alone are not sufficient. Instead, constrained learning with meaningful anatomical region and boundary-aware learning is needed. To address these challenges, we introduce a two stage nnUNet based pipeline that incorporates LA anatomy into the learning process by segmenting the LA cavity in the first stage. Inspired by the observation in \cite{ma2020distance}, where distance maps can enhance segmentation performance, this initial prediction is subsequently used to derive the patient specific wall and cavity signed distance maps (SDMs) to guide scar segmentation. In the second stage, scar segmentation is performed using these geometry-aware representations. The wall constrain and boundary aware uncertainty ROI loss for imbalanced scar learning restricts gradient updates to the atrial wall region, reducing the impact of extreme class imbalance and preventing anatomically implausible scar predictions. The proposed approach enhances scar segmentation performance without modifying the nnUNet architecture, relying instead
on geometry-aware inputs and loss design.
\begin{figure*} [t]
\centering
\includegraphics[width=0.8\linewidth]{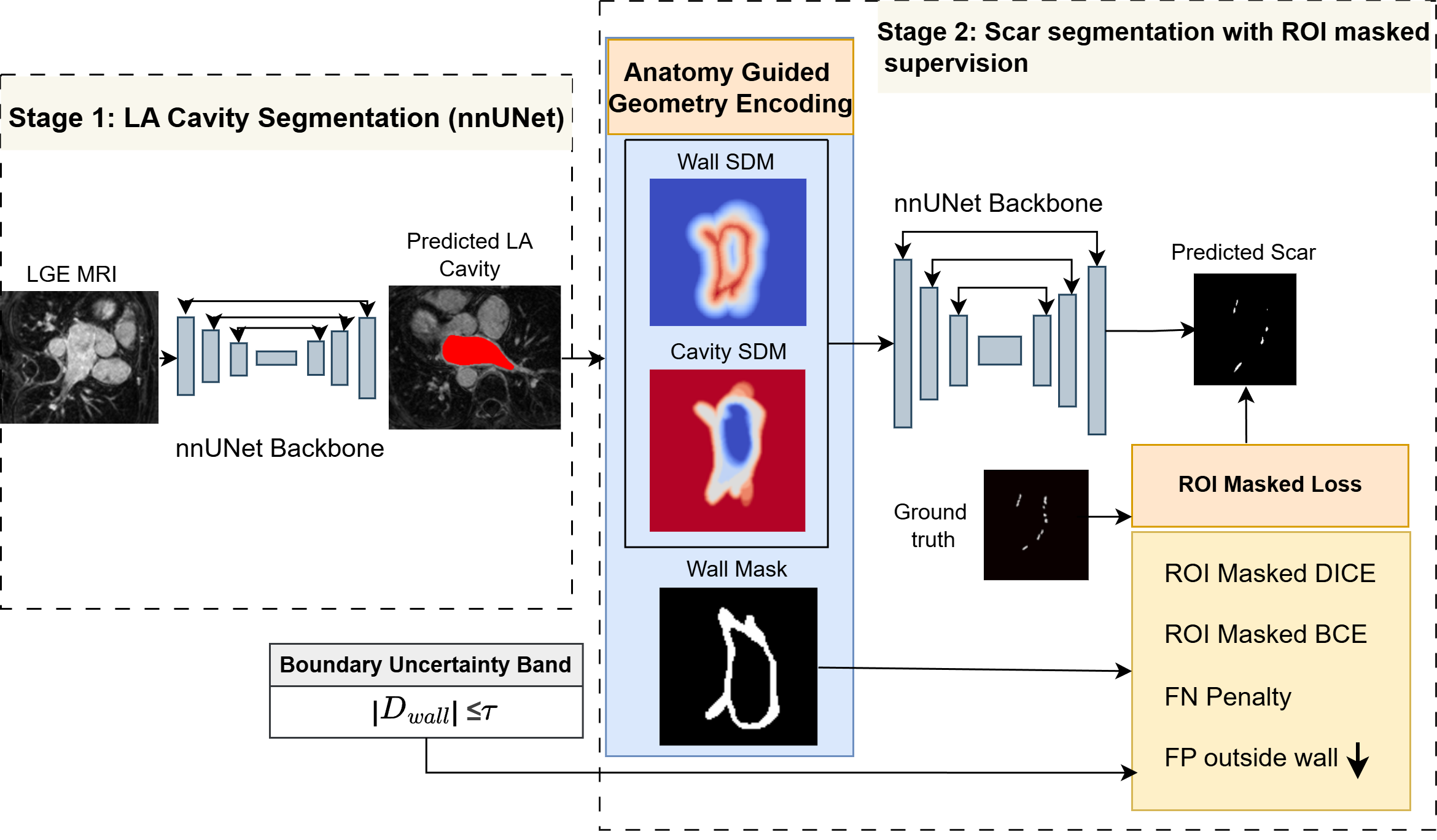}
\caption{Overview of the proposed two-stage anatomy-guided scar segmentation and localization framework.}
  \label{fig1}
  \vspace{-.5em}
\end{figure*}
\section{Methods}
\subsection{Overview}
Motivated by the prior work by Zhou {\it et al.}~\cite{zhou2021probabilistic}, we hypothesize that explicitly incorporating LA anatomical information can improve scar segmentation performance. Since LA cavity segmentation is a well established task, our focus is placed on improving scar segmentation and localization. This task remains challenging due to the very limited publicly available annotated datasets and the inherent challenges imposed by the extremely sparse imbalanced scar regions, and difficulty to to identify thin, low-contrast fibrotic tissue from LGE-MR images.

The backbone of this work is nnUNet-v2\cite{isensee2021nnu}, which we adopt as a strong and established segmentation framework. Fig. \ref{fig1} illustrates the overall pipeline for scar segmentation and localization task. The nnUNet model preserves its original U-shaped framework with an extension to accept multiple input channels, and a customized loss function is implemented via a custom trainer derived from the standard nnUNet trainer class. In addition, boundary uncertainty aware supervision is incorporated to improve robustness in anatomically ambiguous regions. 

\subsection{Dataset and Implementation }
All experiments were conducted on 3D LGE MRI from the LAScarQS 2022 (Task 1, Task 2) dataset\cite{Lees-Miller-LaTeX-course-1}. For LA cavity segmentation in stage 1, we used Task 2 as the training set, which contains 130 LGE-MRI volumes with cavity annotations, and trained the model using all available cases. Evaluation was performed on the LA cavity annotations from Task 1 to assess cross-task generalization.

For stage 2 scar segmentation, we used Task 1, comprising 60 LGE-MRI volumes with scar and used 80\% of the data as training and 20\% for testing.
Across both tasks, images had in-plane resolutions of 576 × 576 or 640 × 640, with 44 or 88 slices per volume. For stage 1, we used the default nnUNet configuration and trained the model for 1000 epochs. For stage 2, we employed the proposed multi-channel input representation and customized loss functions, as described in the Methods section, and trained for 350 epochs. All models were implemented in PyTorch and trained on an NVIDIA GeForce RTX 2080 Ti (11 GB) GPU.



\subsection{Distance Map Encoding}
We use the SDMs as a multi-channel input for the 2nd stage. Let $\Omega \subset \mathbb{Z}^3$ denote the voxel grid and $x \in \Omega$ a voxel location.
Let $M_{\mathrm{LA}}:\Omega \rightarrow \{0,1\}$ be the binary LA cavity mask. We define the Euclidean distance transform with voxel spacing
$s=(s_x,s_y,s_z)$ (mm) as
\begin{equation}
D_s(A)(x)=\min_{y\in A}\lVert (x-y)\odot s\rVert_2,
\end{equation}
where $\odot$ denotes element wise multiplication.

The cavity SDM (positive inside the cavity, negative outside) is computed as
\begin{equation}
\mathrm{SDM}_{\text{cav}}(x)
= D_s(\{M_{LA}=1\})(x) - D_s(\{M_{LA}=0\})(x)
\end{equation}
This corresponds to the difference between the distance to the background and the foreground,
yielding $\mathrm{SDM}_{\mathrm{cav}}(x)>0$ inside the LA cavity and $\mathrm{SDM}_{\mathrm{cav}}(x)<0$ outside.

To obtain a symmetric LA wall band around the cavity boundary, we apply a binary dilation and erosion
to $M_{\mathrm{LA}}$ using a structuring element of radius $r$ voxels (chosen to approximate a wall thickness
of $\tau$ = 2 mm). The wall band mask is defined as
\begin{equation}
M_{\mathrm{wall}} (x) \;=\; \delta_r(M_{\mathrm{LA}} (x)\;\oplus\;\varepsilon_r(M_{\mathrm{LA}} (x))
\label{eq:wall_band}
\end{equation}
where $\delta_r(\cdot)$ and $\varepsilon_r(\cdot)$ denote dilation and erosion, respectively, and $\oplus$
denotes the logical XOR for binary mask. Here we define
\begin{equation}
r=\max\!\left(1,\operatorname{round}\!\left(\frac{\tau}{\min(s_x,s_y)}\right)\right)
\end{equation}
using the in plane spacing to make the band thickness stable under through-plane anisotropy.

Using the wall band $M_{\mathrm{wall}}$, the wall SDM is computed analogously as

\begin{equation}
\begin{aligned}
\mathrm{SDM}_{\mathrm{wall}}(x) \;=\; D_s\!\big(\Omega \setminus \{M_{\mathrm{wall}}=1\}\big)(x)\\
\;-\; D_s\!\big(\{M_{\mathrm{wall}}=0\}\big)(x)
\end{aligned}
\label{eq:wall_sdm}
\end{equation}
yielding positive values inside the wall band and negative values outside.

Finally, both distance maps are clipped to a maximum magnitude $c$ and normalized to $[-1,1]$:
\begin{equation}
\widetilde{\mathrm{SDM}}(x) \;=\; \frac{\mathrm{clip}\!\left(\mathrm{SDM}(x),-c,c\right)}{c},
\qquad c=12~\mathrm{mm}
\label{eq:sdm_norm}
\end{equation}




\subsection{Boundary Uncertainty Aware Supervision}

In AFib ablation therapy, fibrotic scar tissues are spatially concentrated near the wall, not in the cavity or surrounding tissue\cite{lefebvre2022lassnet}. However, in LGE-MRI, the LA wall is thin and exhibit poor contrast, and small geometric errors in atrial segmentation can lead to large localization errors near the wall boundary. To explicitly model this uncertainty, we introduce a boundary uncertainty band (BUB) using the SDM of the LA wall.

Let $\Omega \subset \mathbb{R}^3$ denote the image domain, 
$y(x) \in \{0,1\}$  denote the ground truth scar label, and 
$\hat{p}(x) \in [0,1]$ denote the predicted scar probability at voxel $x$.
Let $D_{\mathrm{wall}}(x) \in [-1,1]$ denote the SDM of the LA wall. We define an uncertainty band of width $\tau$ = 3 mm as
\begin{equation}
B_\tau = \left\{ x \in \Omega \;\middle|\; |D_{\mathrm{wall}}(x)| \le \tau \right\}
\end{equation}

The effective supervision region is constructed as the union of the wall mask $R_{\mathrm{wall}}$ and the band:
\begin{equation}
R_{\mathrm{eff}} = R_{\mathrm{wall}} \cup B_\tau
\end{equation}

All loss terms are computed within $R_{\mathrm{eff}}$, 
which increases sensitivity to scar voxels near uncertain wall boundaries while preventing the network from learning spurious scar predictions outside anatomically plausible regions and improves Dice by reducing false negatives (FNs) at the wall boundary.

\subsection{Weighted ROI-Masked Loss function}
Due to the challenges involved in segmenting scar, pixel-wise supervision often leads to boundary hallucination. To avoid label leakage, the anatomical wall ROI masks were used for loss masking and evaluation, and were explicitly removed from the network input to ensure that the model learns from the image intensity.

Let $R(x) \in {0,1}$ denote the binary wall ROI derived from the predicted atrial cavity of stage 1. The ROI-masked training loss is defined as
\begin{equation}
\mathcal{L}_{\mathrm{ROI}}
=
\frac{1}{|R|}
\sum_{x \in \Omega}
R(x)\,\mathcal{L}(\hat{p}(x), y(x)),
\qquad
|R|=\sum_{x \in \Omega}R(x)
\end{equation}

\subsubsection{ROI-masked Dice loss}

The ROI-masked Dice loss optimizes overlap exclusively within the atrial wall:
\begin{equation}
\mathcal{L}^{R}_{\mathrm{Dice}}
=1-\frac{2\sum_x R(x)\hat p(x)y(x)+\epsilon}
{\sum_x R(x)\hat p(x)+\sum_x R(x)y(x)+\epsilon}
\end{equation}
This term reduces boundary hallucination and enforces spatial consistency of scar predictions within the LA wall.
\subsubsection{ROI-masked BCE loss} 

Within the LA wall, scar voxels are extremely sparse. To counter this imbalance, we employ a ROI-masked weighted binary cross-entropy (BCE) loss:
\begin{equation}
\begin{aligned}
\mathcal{L}^{R}_{\mathrm{wBCE}}
=\frac{1}{|R|}\sum_x R(x)\Big[
-w^{+}y(x)\log \sigma(z(x))\\
-(1-y(x))\log(1-\sigma(z(x)))
\Big]
\end{aligned}
\end{equation}
The positive class weight is computed adaptively based on scar prevalence inside the ROI:
\begin{equation}
P=\sum_x R(x)y(x),\qquad
N=\sum_x R(x)(1-y(x)),
\end{equation}
\begin{equation}
w^{+}=\mathrm{clamp}\!\left(\sqrt{\frac{N+\epsilon}{P+\epsilon}},\,1,\,w_{\max}\right),
\end{equation}
where $w_{\max}$= 10. This formulation increases the penalty for false negative predictions when scar is rare i.e. increases the weight, while ROI masking simultaneously suppresses FPs outside the atrial wall. The class weight is clamped to avoid the gradient explosion and the square root scaling moderates weight growth to ensure stable optimization. 

Both Dice and BCE terms are computed exclusively within a wall ROI. This focuses learning on anatomically valid locations and prevents learning driven by background intensity patterns. 

\subsubsection{Combined ROI Loss}
The primary training objective is defined as:
\begin{equation}
\mathcal{L}_{ROI}=\lambda_{\mathrm{Dice}}\mathcal{L}^{R}_{\mathrm{Dice}}
+\lambda_{\mathrm{BCE}}\mathcal{L}^{R}_{\mathrm{wBCE}},
\end{equation}

where  $\lambda_{Dice}$=1 and $\lambda_{BCE}$=2 balance region based overlap and voxel wise classification.
\subsubsection{Final loss function}
To preserve weak global consistency while maintaining anatomically constrained learning, we optionally include a global Dice regularization term:
\begin{equation}
\mathcal{L}=\mathcal{L}_{ROI}+\alpha\,\lambda_{\mathrm{Dice}}\mathcal{L}^{global}_{\mathrm{Dice}},
\end{equation}
where $\alpha << 1 (0.1)$ controls the global supervision and $\mathcal{L}^{global}_{\mathrm{Dice}}$ is computed over the full image domain.

\section{Experiments and Results }

\subsection{Result Analysis}
We primarily evaluated our pipeline using the Dice Similarity Coefficient (DSC) for both LA and scar segmentation. For LA cavity segmentation, the reported DSC was 93.2\%. This accuracy indicates that the geometry priors passed to stage 2 are highly reliable and any residual errors at this level would most likely shift the wall ROI slightly rather than introduce false scar predictions. Therefore, the reported scar performance reflects a conservative estimate. As our primary concern was to segment and localize the scar, we additionally evaluated the localization performance with Average Symmetric Surface Distance (ASSD) \cite{taha2015metrics} and centroid error to provide a comprehensive assessment. The scar segmentation performance is summarized in Table \ref{tab1}, while the scar localization error is reported in Table \ref{tab2}.

\subsubsection{Quantitative segmentation and localization performance}

The baseline nnUNet achieves a mean DSC of 52.4\%$\pm$12.9, which indicates its inherent difficulty with segmenting scar from LGE-MRI. After incorporating anatomical SDMs derived from the predicted LA cavity, the performance improved consistently and increased the DSC to 53.8\%±12.5, reducing the mean ASSD from 2.075 mm to 1.952 mm. Although the absolute ASSD change is small, ASSD shows how well we locate boundaries. Consistent reductions indicate better geometric alignment of scar predictions along the thin atrial wall, which is essential for accurate scar localization. This also shows that geometry-aware representations give useful spatial context for scar segmentation.

Further gains are observed after introducing ROI  (wall ROI) supervision, which restricts learning to anatomically plausible wall regions. The nnUNet with Anatomy SDM and ROI Loss configuration improves the DSC to 58.4\% ± 11.7 and further reduces the centroid error and ASSD, highlighting that ROI loss is reducing FP predictions outside the atrial wall. Finally, the model combining the anatomy SDMs, ROI loss with BUB achieves the best overall performance, with a DSC of 61.1\% ± 11.5 and the lowest centroid error (4.84±2.4 mm). These results indicate improved spatial localization of scar tissue along the atrial wall, beyond voxel-wise overlap. The DSC was compared with nnUNet and nnUNet with SDM+ROI, the proposed method improves DSC on the test set, with statistically significant differences (paired t-test, p $<$ 0.05 for both comparisons).

\begin{table}[t]
\caption{Segmentation evaluation, mean scores (\%) $\pm$ SD for scar segmentation in the LaScarQS dataset, * p $<$ 0.05 indicates statistical significance}
\vspace{-0.8em}
\label{tab1}
\centering
\setlength{\tabcolsep}{6pt}

\begin{tabularx}{\linewidth}{|Y|c|c|c|}
\hline
\textbf{Method} & \textbf{DSC (\%)} & \textbf{Centroid Error} & \textbf{ASSD (mm)} \\
\hline
nnUNet
& \Tstrut 52.4 $\pm$ 12.9 \Bstrut
& \Tstrut 6.6 $\pm$ 2.6 \Bstrut
& \Tstrut 2.075 $\pm$ 1.3 \Bstrut \\
\hline
nnUNet (SDM)
& \Tstrut 53.8 $\pm$ 12.5 \Bstrut
& \Tstrut 5.4 $\pm$ 2.2 \Bstrut
& \Tstrut 1.952 $\pm$ 1.2 \Bstrut \\
\hline
nnUNet (SDM, ROI)
& \Tstrut 58.4 $\pm$ 11.7 \Bstrut
& \Tstrut 5.1 $\pm$ 2.5 \Bstrut
& \Tstrut 1.732 $\pm$ 1.0 \Bstrut \\
\hline
\textbf{nnUNet (SDM, BUB, ROI*)}
& \textbf{\Tstrut\vphantom{\begin{tabular}{c}0\\0\end{tabular}} 61.1 $\pm$ 11.5* \Bstrut}
& \textbf{\Tstrut\vphantom{\begin{tabular}{c}0\\0\end{tabular}} 4.84 $\pm$ 2.4 \Bstrut}
& \textbf{\Tstrut\vphantom{\begin{tabular}{c}0\\0\end{tabular}} 1.711 $\pm$ 1.2 \Bstrut} \\
\hline
\end{tabularx}
\vspace{-0.8em}
\end{table}

\subsubsection{Anatomical Error Analysis}
We further evaluate the anatomical consistency to validate the localization of scar. Table \ref{tab2} reports FP and FN rates within the relevant regions. The baseline nnUNet produces a percentages of FP predictions outside the atrial wall (22.5\% ± 6.9), reflecting how intensity based models hallucinate scar in non-anatomical regions. These reductions are clinically significant, as FP predictions outside the atrial wall may mislead ablation planning, while FN predictions along the wall may result in missed arrhythmical substrate. While the inclusion of anatomy SDMs provides a modest reduction in FP outside the wall, anatomically implausible predictions remain prevalent without explicit region constraints.


\begin{table}[t]
\caption{Anatomical error analysis (\%) of atrial scar segmentation in the LaScarQS dataset.}
\vspace{-0.8em}
\label{tab2}
\centering
\setlength{\tabcolsep}{6pt}
\renewcommand{\arraystretch}{1.3}

\begin{tabularx}{\linewidth}{|Y|c|c|c|}
\hline
\textbf{Method} & \textbf{FP in cavity} & \textbf{FP outside wall} & \textbf{FN inside wall} \\
\hline
nnUNet 
& \Tstrut 7.2 $\pm$ 5.2 \Bstrut 
& \Tstrut 22.5 $\pm$ 6.9 \Bstrut 
& \Tstrut 20.7 $\pm$ 4.7 \Bstrut \\
\hline
nnUNet (SDM) 
& \Tstrut 6.9 $\pm$ 5.5 \Bstrut 
& \Tstrut 22.1 $\pm$ 6.5 \Bstrut 
& \Tstrut 20.2 $\pm$ 5.4 \Bstrut \\
\hline
nnUNet (SDM, ROI)
& \Tstrut 6.6 $\pm$ 5.1 \Bstrut 
& \Tstrut 21.7 $\pm$ 6.7 \Bstrut 
& \Tstrut 19.3 $\pm$ 5.2 \Bstrut \\
\hline
nnUNet (SDM, BUB, ROI)
& \textbf{\Tstrut\vphantom{\begin{tabular}{c}0\\0\end{tabular}} 6.5 $\pm$ 5.2 \Bstrut}
& \textbf{\Tstrut\vphantom{\begin{tabular}{c}0\\0\end{tabular}} 19.5 $\pm$ 7.7 \Bstrut}
& \textbf{\Tstrut\vphantom{\begin{tabular}{c}0\\0\end{tabular}} 18.6 $\pm$ 4.8 \Bstrut} \\
\hline

\end{tabularx}
\vspace{-0.8em}
\end{table}

The introduction of ROI-masked supervision reduces FP prediction rates outside the atrial wall, decreasing FP outside the wall to 21.7\% ± 6.7, while also reducing FN predictions inside the wall. The full model with boundary uncertainty further improves anatomical consistency, achieving the lowest FP rate outside the wall (19.5\% ± 7.7) and the lowest FN rate inside the wall (18.6\% ± 4.8). This indicates that boundary-uncertainty-aware supervision effectively recovers true scar voxels near ambiguous wall boundaries while maintaining strict anatomical constraints elsewhere.
\begin{figure} [!ht] 
\centering
\includegraphics[width=3.5in]{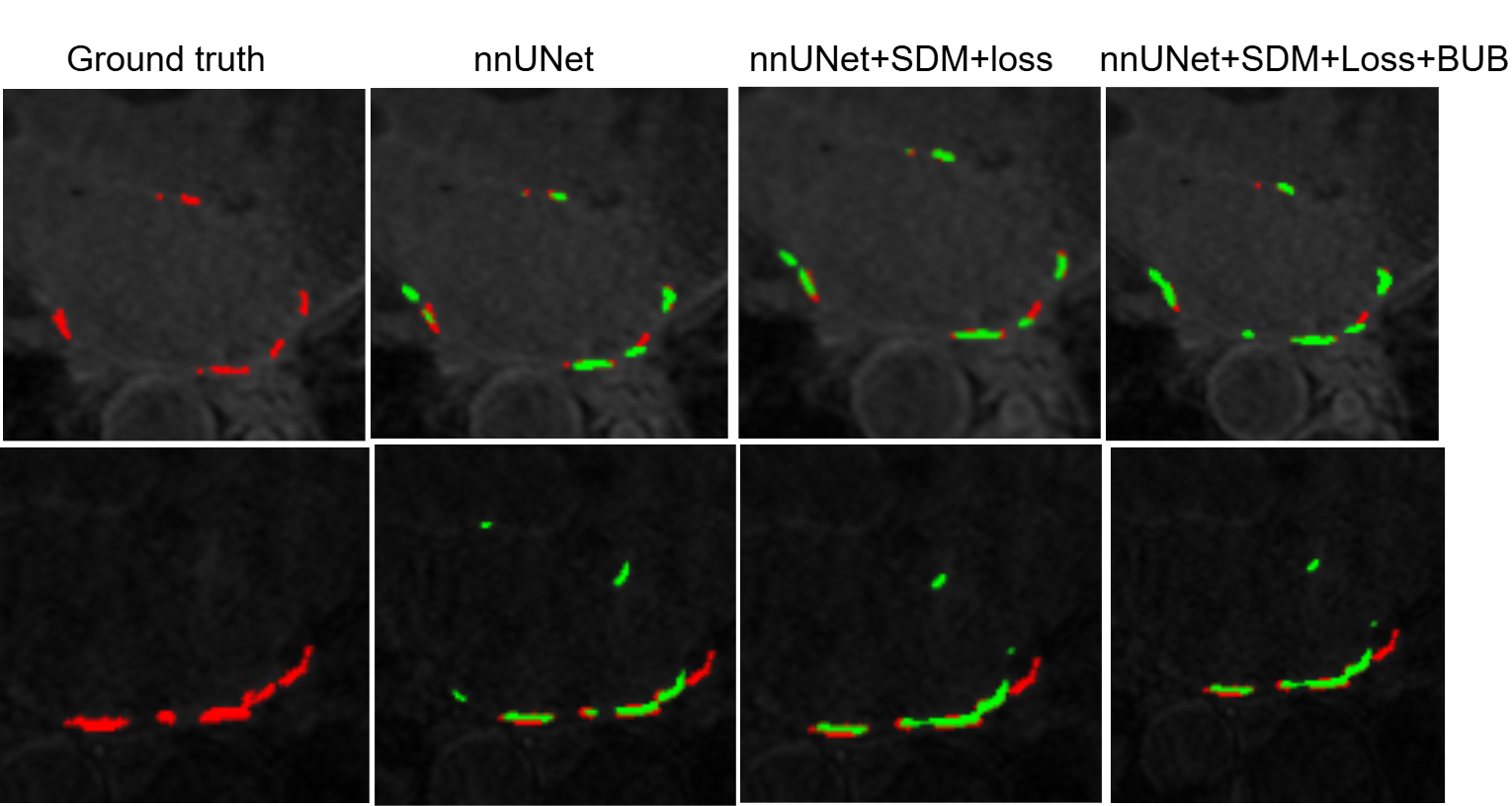}
\caption{Axial views illustrating overlaid predictions (green) and 
ground truth (red) scar localization WITH LA anatomy across different model variants.}
\label{fig2}
\end{figure}
\subsubsection{Qualitative Analysis}
We provide a qualitative comparison  of scar localization across different model variants in Fig. \ref{fig2}. The baseline nnUNet exhibits fragmented scar predictions and produces FPs outside the atrial wall. Incorporating anatomy-guided SDMs improves spatial alignment of predicted scars with the atrial wall but still shows gaps and inconsistent boundary coverage. In contrast, the SDMs with boundary-uncertainty-aware supervision yields more continuous scar delineation along the atrial wall. Although there are discontinuities, these are expected given the weak contrast and partial volume effects in LGE-MRI, particularly in regions of very thin atrial wall. Overall, these observations are consistent with the quantitative reductions in centroid error and FP predictive rates reported in both Tables.



\section{CONCLUSIONS}

This study presents a framework for segmenting and localizing scar from LGE-MRI. By leveraging a reliable LA cavity segmentation in the first stage, we derived the patient specific cavity and wall SDMs to guide scar learning in the second stage. The boundary uncertainty-aware supervision and ROI-masked weighted loss limit learning to anatomically plausible areas. While the test set is limited, we report consistent improvements across all ablation settings. This approach reduces boundary ambiguity and extreme class imbalance. Experimental results highlight improved scar Dice, reduced surface errors, and fewer anatomically implausible FP predictive rates compared to baseline nnUNet configurations. In addition to these performance improvements, this work demonstrates that enforcing anatomical validity through geometry-aware representations and controlled supervision is more effective than architectural complexity for atrial scar segmentation. This paradigm offers a methodological foundation for clinically reliable cardiac image analysis.






\section*{ACKNOWLEDGMENTS}
We would like to acknowledge the generous support for this work by the National Institutes of Health – National Institute of General Medical Sciences under Award No. R35GM128877 and the National Science Foundation - Division of Chemical, Bioengineering and Transport Systems under Award No. 2245152. We also acknowledge the use of ChatGPT\cite{OpenAI2024ChatGPT} for minor assistance with code debugging and clean-up, as well as improving language clarity.

\bibliographystyle{ieeetr}
\bibliography{references}

\begin{thebibliography}{10}

\bibitem{AbbieJordan}
{Science News}, ``Rates of atrial fibrillation are speeding up: What can we do?.'' 2024, {https://www.sciencefocus.com/news/this-heart-condition-is-three-times-more-prevalent-in-the-us-than-previously-thought-heres-how-to-spot-it}.
\newblock (Accessed: 20 January 2026).

\bibitem{recur-rate}
M.~Clinic, ``When atrial fibrillation (afib) ablation timing impacts recurrence?.'' 2025, {https://www.mayoclinic.org/medical-professionals/cardiovascular-diseases/news/when-atrial-fibrillation-afib-ablation-timing-impacts-afib-recurrence/mac-20584500}.
\newblock (Accessed: 20 January 2025).

\bibitem{Marrouche2014}
N.~F. e.~a. Marrouche, ``Association of atrial tissue fibrosis identified by delayed enhancement mri and atrial fibrillation catheter ablation,'' {\em JAMA}, vol.~311, no.~5, pp.~498--506, 2014.

\bibitem{zhou2022edge}
S.~Zhou, K.-N. Wang, and G.-Q. Zhou, ``Edge-enhanced feature guided joint segmentation of left atrial and scars in lge mri images,'' in {\em Challenge on Left Atrial and Scar Quantification and Segmentation}, pp.~93--105, Springer, 2022.

\bibitem{kundu2025investigating}
B.~Kundu, B.~Khanal, R.~Simon, and C.~A. Linte, ``Investigating the domain adaptability of general-purpose foundation models for left atrium segmentation from mr images,'' in {\em International Conference on Functional Imaging and Modeling of the Heart}, pp.~275--287, Springer, 2025.

\bibitem{kundu2025multi}
B.~Kundu, Z.~Yang, R.~Simon, and C.~Linte, ``Multi-scale feature fusion with image-driven spatial integration for left atrium segmentation from cardiac mr images,'' in {\em 2025 47th Annual International Conference of the IEEE Engineering in Medicine and Biology Society (EMBC)}, pp.~1--4, IEEE, 2025.

\bibitem{yang2017fully}
G.~Yang, X.~Zhuang, H.~Khan, S.~Haldar, E.~Nyktari, X.~Ye, G.~Slabaugh, T.~Wong, R.~Mohiaddin, J.~Keegan, {\em et~al.}, ``A fully automatic deep learning method for atrial scarring segmentation from late gadolinium-enhanced mri images,'' in {\em 2017 IEEE 14th International Symposium on Biomedical Imaging (ISBI 2017)}, pp.~844--848, IEEE, 2017.

\bibitem{khan2022sequential}
A.~Khan, O.~Alwazzan, M.~Benning, and G.~Slabaugh, ``Sequential segmentation of the left atrium and atrial scars using a multi-scale weight sharing network and boundary-based processing,'' in {\em Challenge on Left Atrial and Scar Quantification and Segmentation}, pp.~69--82, Springer, 2022.

\bibitem{zhang2024left}
Y.~Zhang, H.~Cheng, D.~Li, and L.~Pan, ``Left atrial scar segmentation and quantification using residual cbam-eam attention unet for lge mri,'' in {\em MICCAI Challenge on Comprehensive Analysis and Computing of Real-World Medical Images}, pp.~149--157, Springer, 2024.

\bibitem{li2022atrialjsqnet}
L.~Li, V.~A. Zimmer, J.~A. Schnabel, and X.~Zhuang, ``Atrialjsqnet: a new framework for joint segmentation and quantification of left atrium and scars incorporating spatial and shape information,'' {\em Medical image analysis}, vol.~76, p.~102303, 2022.

\bibitem{lefebvre2022lassnet}
A.~L. Lefebvre, C.~A. Yamamoto, J.~K. Shade, R.~P. Bradley, R.~A. Yu, R.~L. Ali, D.~M. Popescu, A.~Prakosa, E.~G. Kholmovski, and N.~A. Trayanova, ``Lassnet: a four steps deep neural network for left atrial segmentation and scar quantification,'' in {\em Challenge on Left Atrial and Scar Quantification and Segmentation}, pp.~1--15, Springer, 2022.

\bibitem{Lees-Miller-LaTeX-course-1}
LaScarQS, ``Left atrial and scar quantification \& segmentation challenge.'' 2022, {https://zmiclab.github.io/projects/lascarqs22}.
\newblock (Accessed: 20 January 2025).

\bibitem{ma2020distance}
J.~Ma, Z.~Wei, Y.~Zhang, Y.~Wang, R.~Lv, C.~Zhu, C.~Gaoxiang, J.~Liu, C.~Peng, L.~Wang, {\em et~al.}, ``How distance transform maps boost segmentation cnns: an empirical study,'' in {\em Medical Imaging with Deep Learning}, pp.~479--492, PMLR, 2020.

\bibitem{zhou2021probabilistic}
X.~Zhou, V.~Koltun, and P.~Kr{\"a}henb{\"u}hl, ``Probabilistic two-stage detection,'' {\em arXiv preprint arXiv:2103.07461}, 2021.

\bibitem{isensee2021nnu}
F.~Isensee, P.~F. Jaeger, S.~A. Kohl, J.~Petersen, and K.~H. Maier-Hein, ``nnu-net: a self-configuring method for deep learning-based biomedical image segmentation,'' {\em Nature methods}, vol.~18, no.~2, pp.~203--211, 2021.

\bibitem{taha2015metrics}
A.~A. Taha and A.~Hanbury, ``Metrics for evaluating 3d medical image segmentation: analysis, selection, and tool,'' {\em BMC Medical Imaging}, vol.~15, no.~1, p.~29, 2015.

\bibitem{OpenAI2024ChatGPT}
{OpenAI}, ``Chatgpt.'' \url{https://chat.openai.com}, 2024.
\newblock Large language model accessed January 2026.

\end{thebibliography}

\end{document}